# Study of Thick CZT Detectors for X-ray and Gamma-Ray Astronomy

Qiang Li[1,4], M. Beilicke[1], Kuen Lee[1], Alfred Garson III[1], Q. Guo[1], Jerrad Martin[1], Y. Yin[1], P. Dowkontt[1], G. De Geronimo[2], I. Jung[3], H. Krawczynski[1]
1. Washington University in St. Louis, 2. Brookhaven National Laboratory, 3. Universität Erlangen-Nürnberg, 4. Present at State Key Laboratory of Solidification Processing, School of Materials Science and Engineering, Northwestern Polytechnical University, Xi'an 710072, China

**Abstract:**

CdZnTe (CZT) is a wide bandgap II-VI semiconductor developed for the spectroscopic detection of X-rays and γ-rays at room temperature. The Swift Burst Alert Telescope is using an 5240 cm$^2$ array of 2 mm thick CZT detectors for the detection of 15-150 keV X-rays from Gamma-Ray Bursts. We report on the systematic tests of thicker (≥0.5 cm) CZT detectors with volumes between 2 cm$^3$ and 4 cm$^3$ which are potential detector choices for a number of future X-ray telescopes that operate in the 10 keV to a few MeV energy range. The detectors contacted in our laboratory achieve Full Width Half Maximum energy resolutions of 2.7 keV (4.5%) at 59 keV, 3 keV (2.5%) at 122 keV and 4 keV (0.6%) at 662 keV. The 59 keV and 122 keV energy resolutions are among the world-best results for ≥0.5 cm thick CZT detectors. We use the data set to study trends of how the energy resolution depends on the detector thickness and on the pixel pitch. Unfortunately, we do not find clear trends, indicating that even for the extremely good energy resolutions reported here, the achievable energy resolutions are largely determined by the properties of individual crystals. Somewhat surprisingly, we achieve the reported results without applying a correction of the anode signals for the depth of the interaction. Measuring the interaction depths thus does not seem to be a pre-requisite for achieving sub-1% energy resolutions at 662 keV.

## 1. Introduction

Over the last two decades, the II-VI semiconductor CdZnTe (CZT) has emerged as the material of choice for room temperature detection of hard X-rays and soft γ-rays. The techniques of growing the crystals, the design of the detectors, and the electronics used for reading out the detectors have been considerably improved over the last few years [1-3]. The material finds applications in astrophysics and particle physics experiments, and in medical and homeland security applications.

An example of a high-energy astrophysics experiment is the Burst Alert Telescope (BAT) on board of the Swift UV-X-ray observatory, launched on Nov. 20, 2004. The experiment uses an array of 32,768 CZT detectors (each: 0.2×0.4×0.4 cm$^3$) to detect X-rays in the 15 -150 keV energy range [4]. A number of proposed future experiments will use thicker (≥0.5 cm) CZT detectors. The EXIST (Energetic X-ray Imaging Survey Telescope) mission is a proposed all sky survey hard X-ray survey telescope that would use ~11,250 CZT detectors (each: 0.5×2.0×2.0 cm$^3$) in a coded mask imager with a detector area of 4.5 m$^2$ [5]. HX-POL is a proposed hard X-ray polarimeter which would combine a stack of thick Si detectors as low-Z Compton scatter medium and CZT detectors (each: 0.5×3.9×3.9 cm$^3$) as high-Z photoeffect absorber [6].

The Cadmium-Zinc-Telluride 0-neutrino Double-Beta Decay Apparatus (COBRA) is a particle physics experiment that uses CZT both as a source and a detector to investigate the mass of the neutrino [7]. A large-scale COBRA experiment would be made of 420 kg of CZT detectors and use either coplanar grid detectors ($1\times1\times1$ cm$^3$), or pixelated detectors ($0.5\times3.9\times3.9$ cm$^3$) [8]. In the field of medical applications, CZT detectors are an interesting option for Positron Emission Tomography (PET) and Single Photon Emission Computed Tomography (SPECT). Homeland security applications include the development of coded mask and Compton imaging detectors for detection, localization, and identification of nuclear threats at our borders [9,10].

CZT detectors have good electron mobility-lifetime products ($\mu_e\tau_e \sim 10^{-3}$-$10^{-2}$ cm$^2$ V$^{-1}$), but poor hole mobility-lifetime products ($\mu_h\tau_h \sim 5\times10^{-5}$ cm$^2$ V$^{-1}$). All state-of-the-art CZT detectors use "single polarity" readout schemes, where the main information about the energy of the detected radiation is inferred from the anode signals. Coplanar grid detectors, pixelated detectors, and Frisch grid detectors can overcome the severe hole trapping problem and greatly improve the energy spectra resolution of large-volume CZT detectors. Zhang et al. [11] tested CZT detectors made from High Pressure Bridgeman (HPB) CZT. For one, especially good $1.5\times2\times2$ cm$^3$ CZT detector ($11\times11$ pixels, pixel pitch: 1.8 mm), a 662 keV energy resolution of 3.3 keV (0.5%) FWHM was reported after correcting the signals for the 3-D position of the interaction. The test of a larger sample of detectors revealed poorer "typical" energy resolutions of between 1% (7 keV) and 2% (13 keV) [12].

Over the last few years, our group studied CZT detectors grown with the modified horizontal Bridgman (MHB) method by the company Orbotech Medical Solutions Ltd. In [13] we reported results obtained when combining a monolithic Au cathode with different anode contacts made of metals. Four different metals were used for the anode: Indium, Titanium, Chromium and Gold with work-functions between 4.1 eV and 5.1 eV. The best performances were achieved with anode contacts made of the low-work-function metals Indium and Titanium and a cathode made of the high-work-function metal Au. At lower ambient temperatures (down to -30°C), the detector performance deteriorates slightly, an effect that can be compensated by increasing the bias voltage [14].

Studies of the response of ≥0.5 cm thick CZT detectors with collimated X-ray beams can be found in [15-17].

In this paper, we present a systematic study of the energy resolution and detection efficiency of CZT detectors made from both, MHB and HPB CZT. The tests use a low-noise readout ASIC developed at Brookhaven National Laboratory. We evaluate the performance of substrates of different thicknesses when contacted with pixels at different pitches. Each substrate is used multiple times with different pixel patterns by polishing off one pattern and depositing a new pattern. This procedure has the advantage that the performance of different pixel patterns can be assessed largely free from sample-to-sample variations.

The rest of the paper is organized as follows. The detector fabrication methods are described in Sect. 2. The ASIC based readout system and the detector mounting are described in Sect. 3. The results obtained with the different CZT detectors contacted with different pixel patterns are presented in Sect. 4. A summary and a discussion are given in Sect. 5.

In the following all energy resolutions are given as Full Width half Maximum (FWHM)

values. We use $^{241}$Am (59.5 keV), $^{57}$Co (122 keV, 136 keV), and $^{137}$Cs (662 keV) as X-ray and gamma-ray sources.

## 2. Detector fabrication

The studies use MHB CZT from the company Orbotech Medical Solutions and HPB CZT from the company eV-Product. We used 0.5 cm, 0.75 cm and 1 cm thick crystals with a 2×2 cm$^2$ footprint. The Orbotech detectors were delivered with a planar In cathode and 8×8 In anode pixels. After evaluating the detector performance, we replaced the In anode with an Au cathode which lead to an improvement of the detector performance for all tested crystals. The eV-Product detectors were delivered with a planar Pt cathode and 8×8 Pt pixels. After testing the detectors with the delivered pixels, we replaced the pixels with Ti pixels at different pixel pitches.

The detector fabrication starts with removing the old pixel patterns by polishing with different grades of abrasive paper and alumina suspension with particles sizes down to 0.05 μm. We found that etching with a 5%-95% Br-Methanol solution does not only improve the electrical properties of the contacts, but also improved the adhesion of the contacts. After surface preparation, a standard photolithographic process is used consisting of applying the photoresist Shipley S1813, pre-baking, exposure, post-baking, and development with the developer Shipley Cd-30. A metal film is then deposited on both the exposed and photoresist-covered surface of the CZT with an electron beam evaporator. Finally, the remaining photoresist is removed with acetone, which also takes away the metal film deposited on the photoresist. The parameters from a systematic optimization of the photolithography process were used [18].

The parameters of the pixel patterns used in this paper are given in Table 1.

**Table 1:** Parameters of tested pixels patterns.

| Number of pixels | Pixel Pitch [cm] | Pixel Width [cm] | Gap Between Pixels [cm] |
|---|---|---|---|
| 8×8 | 0.25 | 0.16 | 0.09 |
| 11×11 | 0.172 | 0.157 | 0.015 |
| 15×15 | 0.13 | 0.1 | 0.03 |

## 3. Signal amplification and readout

### 3.1 The NCI-ASIC developed at Brookhaven National Laboratory

The results presented in this paper are from data acquired with a readout system based on the "NCI-ASIC" developed by Brookhaven National Lab and NRL [19]. The ASIC die has a footprint of 4.922×4.922 mm$^2$ and is fabricated in CMOS 0.25μm technology. It combines excellent noise performance (~1 keV Si) with a large dynamic range (>100) and low power dissipation (4 mW/channel + 38 mW/ASIC). For each of its 32 channels, the ASIC provides a low-noise preamplifier [20-22], a fifth order filter (shaper) with baseline stabilizer [23], a threshold comparator, and a peak detector with analog memory [24]. The ASIC properly processes charges of either polarity by using a design with low-noise continuous reset circuits for each polarity [25, 26]. The NCI-ASIC offers programmable shaping time (0.5us, 1us, 2us, and 4us) and gain (14.25 mV/fC 28.5 mV/fC, and 57 mV/fC). At the lowest gain, the dynamic range of the ASIC extends to 30 fC (~4 MeV in CZT). The measurements presented in this paper are taken with the low or medium gain setting. The read-out is triggered by a single threshold discriminator which can be set globally (step size of 2mV) for all channels. An additional channel trim allows to adjust the threshold for individual channels within a 4-bit dynamic range of 56mV, relative to the global threshold. Figure 1 shows the readout board developed at Washington University. We have recently developed scalable version of the readout system that allows us to read out large arrays of CZT detectors. This latter system will be described in a forthcoming paper.

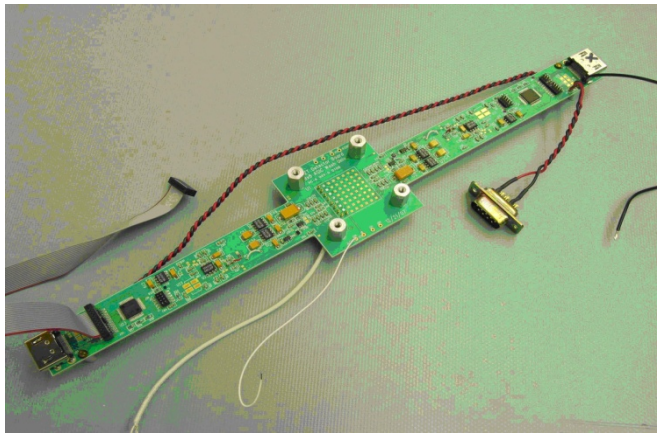

Figure 1 Board for the readout of pixilated detectors developed at Washington University. The board uses two NCI-ASICs developed at Brookhaven National Laboratory to read out the negative polarity anode signals. The two ASICs are mounted at the "bottom" of the board and cannot be seen in the image. At the center of the board, contact pads for connecting the board to a 64-pixel detector can be seen. A third ASIC on a similar board (not shown here) is used to read out the positive polarity cathode signal.

The data acquisition mode of the ASIC continues for three micro seconds after the first channel triggers so that multi-channel events have sufficient time to settle. Subsequently, the data of the triggered channels (pulse height) is transferred through an I/O board to a computer for data processing and storage.

The ASIC features a 200 fF test capacitor to inject charges into individual channels. The charge injection data can be used to test the read-out noise and the linearity of individual channels. The results of the internal test pulser were validated with an external pulser with a known charge output. The results obtained with the internal test capacitor and the external charge injection agree with the exception of the highest signal amplitudes at which the internal test circuit saturates.

For each ASIC, we use the following test program: (i) Channel by channel, test pulses of fixed charge are injected and the measured signal amplitudes are read out; the histogrammed results are used to determine the electronic noise and to identify problematic channels. (ii) The test pulses are again injected, however amplitude and gain settings are varied (Fig. 2). For each combination of pulse amplitude and gain setting, the results are histogrammed and fitted with a Gaussian. The fit parameters are used to characterize the channel responses, i.e. to determine the channel gains, pedestals, and linearity. The measurements are performed twice, before and after mounting the CZT detector. The procedure allows us to calibrate data and find potential problems with the ASIC and with the CZT detector. After the charge injection calibration measurements, the amplitude-to-keV conversion factors are determined by irradiating the detector with a $^{137}$Cs source and measuring the position of the photopeak for each channel. We operate at 18˚C and the shaping time is 1 μsec.

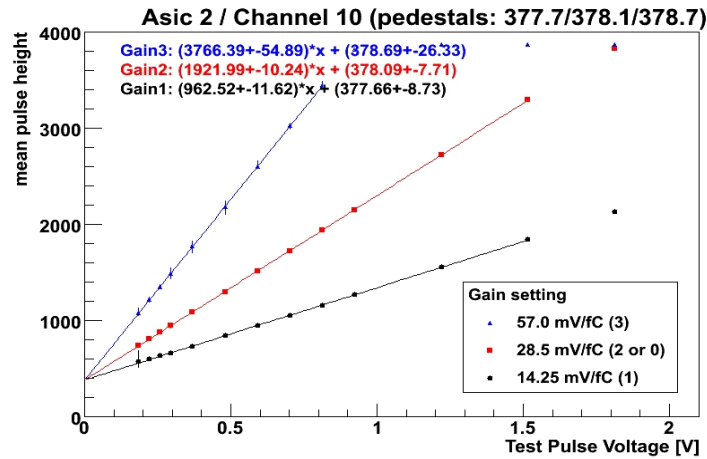

Figure 2: The plot shows the linearity characteristics of an individual ASIC channel for the three different gain settings (high gain:blue, medium gain:red, low gain:black). The test pulse voltage amplitude is increased (X axis) and the pulse height channels are recorded (Y axis) for each amplitude. The pulse height distributions are fit with linear functions. The fit parameters are used to determine the gain and the pedestal of the channel. The higher gain settings reach a pulse height plateau at given test pulse aplitudes above which the output pulse height is constant. This corresponding to the maximum voltage (gain*signal charge) that fits within the ASICs dynamic range.

The different ASIC channels and detector pixels exhibit different noise properties. Minimizing the energy threshold of a measurement for all channels thus requires adjustment the ASIC trigger threshold on a channel-to-channel basis. We developed a procedure to automatically adjusts the threshold settings (global and individual channel trims) to assure that the threshold of each channel is set as low as possible, but still safely above the noise level. As the dynamic range of the channel trims is limited, exceptionally noisy channels are disabled. Using this procedure, we achieve mean energy thresholds of ~30 keV with our CZT detectors.

### 3.2 Mounting of detectors

The CZT detectors are mounted in a holder made of milled Ultem plastic using gold-plated, spring-loaded "pogo-pins" to connect the pixels with readout traces on the PC board. If a pattern has more than 64 pixels, we only read out the central 64 pixels, as our readout system is presently limited to the readout of 64 pixels. Readout systems for detectors with more pixels are under development. For patterns with more than 64 pixels, the other pixels are held at ground potential. The pixel are DC coupled to two 32 channel ASICs programmed to read negative polarity signals. The cathode signal is AC coupled and read out by an additional third ASIC programmed to read positive polarity signals. We use the same readout board to read out 64 pixel detectors and detectors with more pixels. For detectors with more pixels, an adapter board is used to read out the central 64 pixels.

## 4. Results

## 4.1 Energy resolutions and fraction of photopeak events

In the following we present results for two 0.5 cm thick, one 0.75 cm thick, one 1 cm thick Orbotech detectors, and one 1 cm thick eV-Product detector. Initially we had planned to conduct the study with two Orbotech detectors of each thickness. Unfortunately, the ≥0.5 cm detectors are a non-standard product of Orbotech. The study thus uses the custom-grown crystals that were available. The good results obtained with the 0.75 cm and 1 cm thick Orbotech detectors may prompt additional fabrication of such thick crystals. The spectroscopic performance of the detectors was measured by flood illuminating the detectors with $^{57}$Co (1$\mu$Ci) and $^{137}$Cs (1$\mu$Ci), sources. The detectors were biased at -1500~-2500 V at the cathodes (anodes at ground). The photopeak of each spectrum was fit with a superposition of a Gaussian distribution and an exponential tail towards smaller signal amplitudes. For the $^{137}$Cs energy spectra, we define the "fraction of photopeak events" as the ratio of the counts in the Gaussian photopeak (photopeak center ±2σ) to the total counts measured above an energy threshold of 100 keV. The fraction of photopeak events is used to characterize the photopeak detection efficiency of the detectors. The following results do not include any correction for the depth of the interaction (DOI) because for most crystals the correction for the depth of the interaction does not result in a significant improvement of the energy resolution. Corrected energy spectra will be shown in Sect. 4.2.

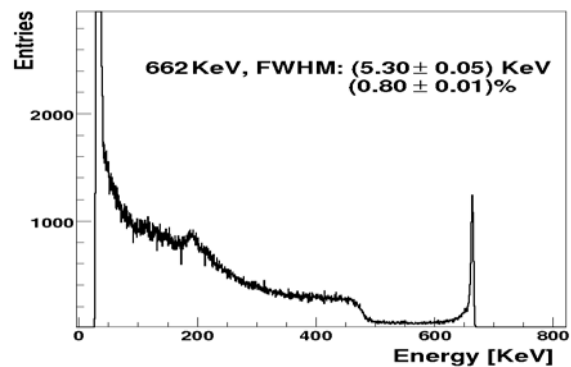

(a)

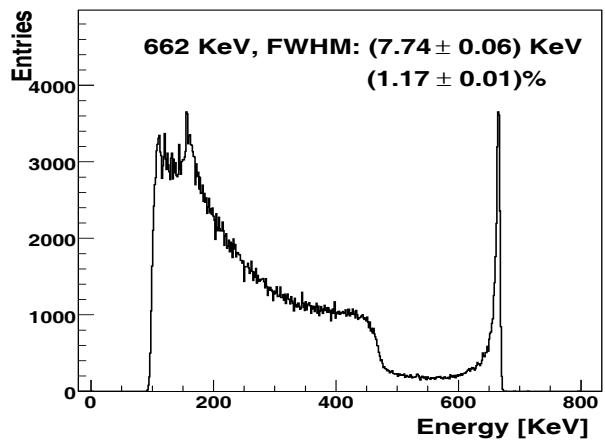

(b)

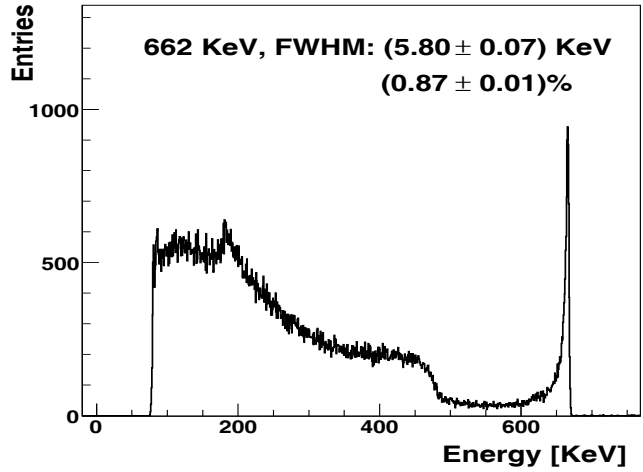

(c)

Figure 3: $^{137}$Cs (662 keV) energy spectra obtained with the 1×2×2 cm$^3$ CZT detector from the company eV-Products. The panels show the results obtained with the 15×15 pixel pattern (a), with the 11×11 pixel pattern (b), and with the 8×8 pixel pattern (c).

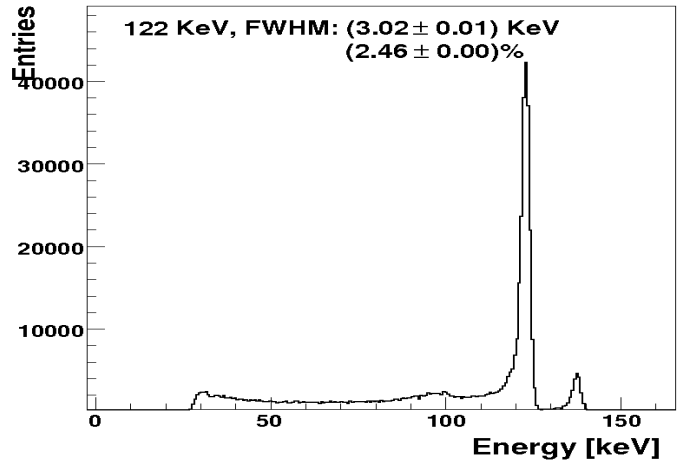

(a)

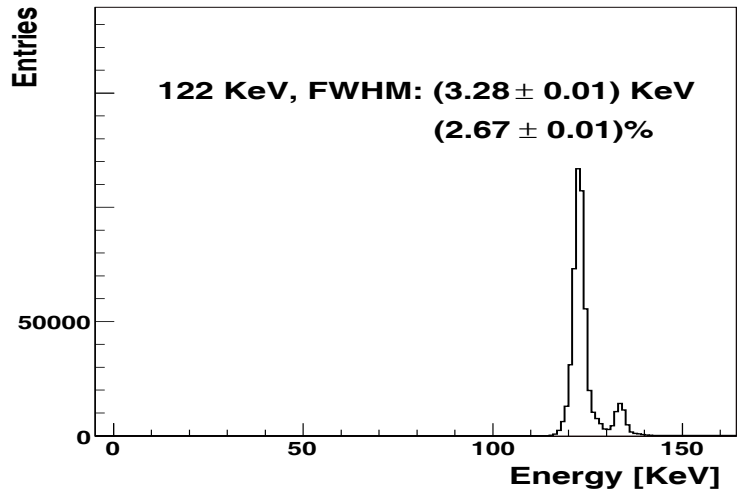

(b)

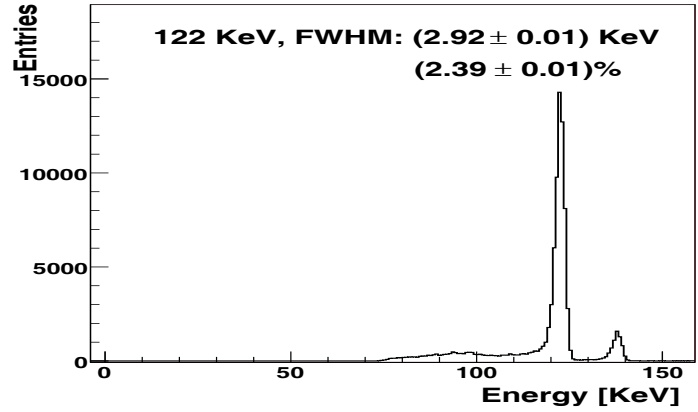

(c)

Figure 4: $^{57}$Co (122 keV and 136 keV) energy spectra obtained with the 1×2×2 cm$^3$ CZT detector from the company eV-Products. The panels show the results obtained with the 15×15 pixel pattern (a), with the 11×11 pixel pattern (b), and with the 8×8 pixel pattern (c).

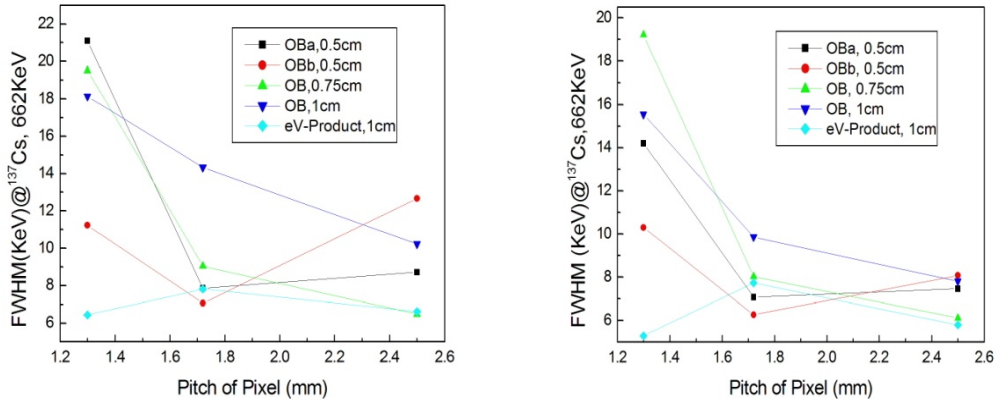

Figure 5: 662 keV ($^{137}$Cs) energy resolution (FWHM) as function of the pixel pitch. The left panel shows the median of all tested pixels. The right panel shows the results for the best pixels. The cathode bias voltages of the detectors range from -2500 V to -1500 V.

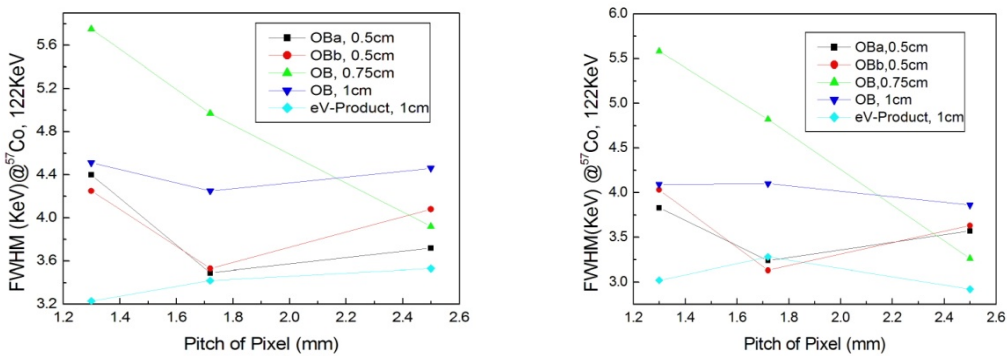

Figure 6: 122 keV ($^{57}$Co) energy resolution (FWHM) as function of the pixel pitch. The left panel shows the median of all tested pixels. The right panel shows the results for the best pixels. The cathode bias voltages of the detectors range from -2500 V to -1500 V.

Exemplary energy spectra are shown in Figs. 3 and 4. At 662 keV, energy resolutions between 0.8% and 3.2% and photopeak fractions between 2% and 8% were measured. At 122 keV, energy resolutions between 2.4%~4.6% were recorded. Figure 5 shows the 662 keV energy resolutions as function of pixel pitch. Figure 6 does the same for the 122 keV energy resolutions.

The results are remarkable in several aspects:

• The energy resolutions reported here are among the best ever observed with ≥0.5 cm thick CZT detectors. Remarkable results include the HPB energy resolutions of 2.9 keV at 122 keV and 5.3 keV at 662 keV and MHB energy resolutions of 3.5 keV at 122 keV and 6.1 keV at 662 keV. The excellent results stem from the low noise of the ASIC based readout system and from our optimization of the contacting process. At 122 keV our results are to our knowledge the best results for thick (≥0.5cm) CZT detectors reported in the literature so far for both HPB and MHB detectors. At 662 keV our results are the best results reported for thick MHB substrates. For results obtained with HPB and MHB CZT detectors by other groups we refer the reader to [27,28].

• The two identical 0.5 cm thick MHB detectors (OBa, OBb) give very different results. The results demonstrate that there are substantial differences between individual substrates.

• The 0.5 cm thick MHB detector OBa shows a rather poor 662 keV energy resolution when contacted with 15x15 pixels. It is not clear where this effect comes from, especially since the same detector with the same pixels performs well at 122 keV.

• The observed variation of the energy resolutions of detectors of the same thickness is larger than the variation of the energy resolutions of detectors of different thicknesses. This fact makes it difficult to detect a trend of how the energy resolution depends on the crystal thickness.

• The energy resolutions obtained with the 1 cm thick HPB crystal do not depend on the used pixel pattern. In contrast, in some cases - but not in all cases – the MHB detectors show a poorer performance when they are read out with pixels at a smaller pixel pitch. The interpretation of the effect is rendered more difficult by the fact that the trend depends on the considered energy.

Unfortunately, the results are not entirely conclusive in terms of the energy resolution as function of the crystal thickness and the pixel pitch. Our main conclusion is that excellent performances can be achieved with a wide range of thicknesses and pixel pitches. Furthermore, the results show that studies limited to single detectors can be misleading. Unfortunately, studies of single detectors are common; for notable exceptions see [12,28-29].

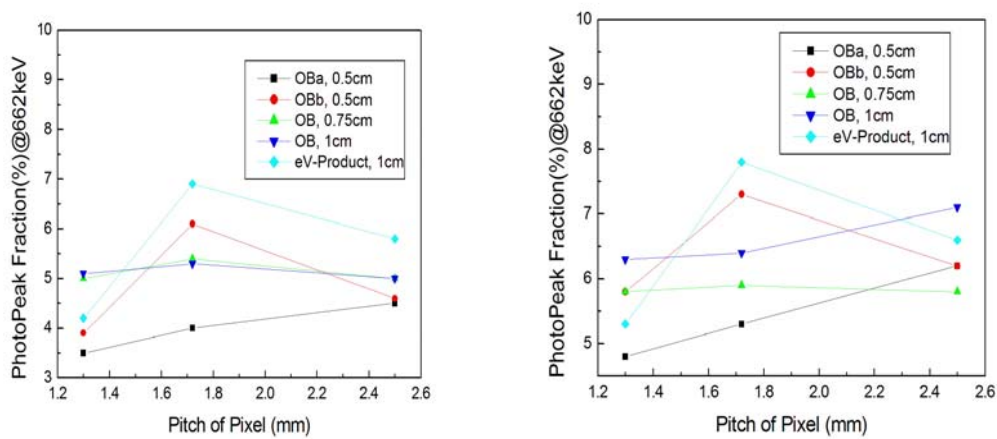

Figure 7: Fraction of photopeak events relative to all >100 keV events determined from flood illuminating the detectors with a $^{137}$Cs source. The left panel shows the median of all tested pixels.

The right panel shows the results for the best pixels. The cathode bias voltages of the detectors range from -2500 V to -1500 V.

Figure 7 shows the fraction of photopeak events as function of pixel pitch measured with the $^{137}$Cs source. The fraction of photopeak events is highest for the 11×11 pixel pattern (change according to the results). Note that this pixel pattern has also the largest ratio of anode area covered by the pixels (rather than by the gaps between pixels).

Tables 2-4 summarize all the results from the $^{137}$Cs and $^{57}$Co measurements.

**Table 2:** Energy resolutions (FWHM, keV) measured at 662 keV ($^{137}$Cs).

| Thickness / Pitch | | 0.5 cm | | 0.75 cm | 1 cm | |
|---|---|---|---|---|---|---|
| | | OBa[1] | OBb | OB | OB | eV[2] |
| 1.3mm (225 pixel) | Best[3] | 14.2±0.3 | 10.3±0.1 | 19.2±0.1 | 15.5±0.1 | 5.3±0.05 |
| | Median[4] | 21.1 | 11.2 | 19.5 | 18.1 | 6.4 |
| | σ[5] | 3.0 | 3.8 | 2.5 | 1.6 | 1.9 |
| 1.72mm (121pixel) | Best | 7.1±0.1 | 6.3±0.1 | 8.0±0.1 | 9.9±0.1 | 7.7±0.06 |
| | Median | 7.9 | 7.1 | 9.0 | 14.3 | 7.8 |
| | σ | 0.9 | 1.4 | 0.4 | 2.2 | 1.2 |
| 2.5mm (64 pixel) | Best | 7.5±0.1 | 8.1±0.1 | 6.1±0.1 | 7.8±0.1 | 5.8±0.1 |
| | Median | 8.7 | 12.7 | 6.5 | 10.2 | 6.6 |
| | σ | 0.9 | 2.8 | 0.9 | 2.0 | 1.2 |

[1] OB denotes an MHB detector from the company Orbotech. [2] eV denotes an HPB detector from the company eV-Products. [3] Result obtained for the best pixel together with statistical errors from the fit of the photopeak with a Gaussian. [4] The median denotes the median resolution of all tested pixels. [5] Width of energy resolution distribution.

**Table 3:** Energy resolutions (FWHM, keV) at 122 keV ($^{57}$Co), see annotations of Table 2.

| Thickness / Pitch | | 0.5 cm | | 0.75cm | 1 cm | |
|---|---|---|---|---|---|---|
| | | OBa | OBb | OB | OB | eV |
| 1.3mm (225 pixel) | Best | 3.8±0.01 | 4.0±0.01 | 5.6±0.02 | 4.1±0.01 | 3.0±0.01 |
| | Median | 4.4 | 4.3 | 5.8 | 4.5 | 3.2 |
| | σ | 0.5 | 0.6 | 0.2 | 0.5 | 0.6 |
| 1.72mm (121 pixel) | Best | 3.2±0.01 | 3.1±0.01 | 4.8±0.01 | 4.1±0.03 | 3.3±0.01 |
| | Median | 3.5 | 3.6 | 5.0 | 4.3 | 3.4 |
| | σ | 1.5 | 2.2 | 1.0 | 1.2 | 0.4 |
| 2.5mm (64 pixel) | Best | 3.6±0.02 | 3.6±0.03 | 3.3±0.01 | 3.9±0.04 | 2.9±0.01 |
| | Median | 3.7 | 4.1 | 3.9 | 4.5 | 3.5 |
| | σ | 0.4 | 1.3 | 0.7 | 0.8 | 1.0 |

**Table 4:** Fraction of photopeak events (% above 100keV) at 662 keV ($^{137}$Cs).

| Thickness / Pitch | 0.5 cm | | 0.75cm | 1 cm | |
|---|---|---|---|---|---|
| | Oba[1] | OBb | OB | OB | eV[2] |

| | | | | | | |
|---|---|---|---|---|---|---|
| 1.3mm (225 pixel) | Best[3] | 4.8 | 5.8 | 5.8 | 6.3 | 5.3 |
| | Median[4] | 3.5 | 3.9 | 5.0 | 5.1 | 4.2 |
| | σ[5] | 0.5 | 0.6 | 0.5 | 0.5 | 0.5 |
| 1.72mm (121 pixel) | Best | 5.3 | 7.3 | 5.9 | 6.4 | 7.8 |
| | Median | 4.0 | 6.1 | 5.4 | 5.3 | 6.9 |
| | σ | 0.5 | 1.2 | 0.5 | 0.8 | 3.7 |
| 2.5mm (64 pixel) | Best | 6.2 | 6.2 | 5.8 | 7.1 | 6.6 |
| | Median | 4.5 | 4.6 | 5.0 | 5.0 | 5.8 |
| | σ | 0.9 | 0.5 | 0.5 | 0.9 | 0.5 |

[1] OB denotes an MHB detector from the company Orbotech. [2] eV denotes an HPB detector from the company eV-Products. [3] Result obtained for the best pixel. [4] The median denotes the median resolution of all tested pixels. [5] Width of distribution.

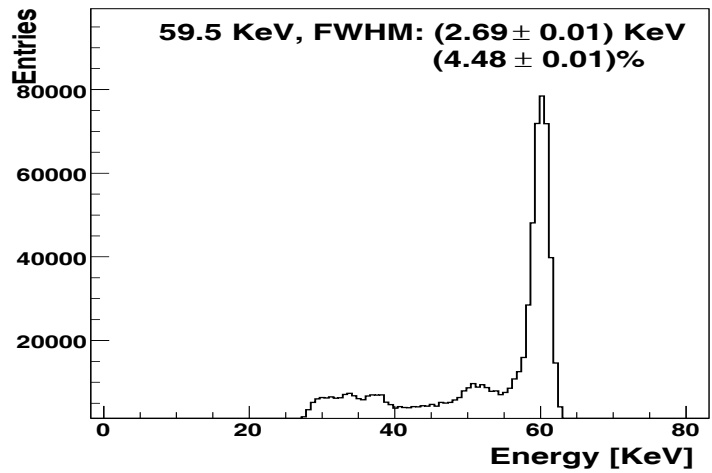

Figure 8: $^{241}$Am (59.5 keV) energy spectra obtained with the 1×2×2 cm$^3$ CZT detector from the company eV-Products. The energy resolution is 2.69 keV (4.48%) for the best pixel.

Towards the end of the study (which took ~1 year), we implemented a new algorithm that allowed us to lower the energy threshold from ~80 keV to 20 keV. For this purpose, a few channels (typically 1 or 2) had to be disabled, because they trigger with very high rates at lower threshold settings. As an example of an energy spectrum taken with a low energy threshold, Fig. 8 shows a $^{241}$Am (59.5 keV) energy spectrum taken with the 1×2×2 cm$^3$ HPB detector from the company eV-Products. The 59 keV energy resolutions are 4.48% (2.69 keV) for the best pixels. The median values are 4.87% (2.92 keV).

## 4.2 Depth of interaction correction

Most current CZT detectors use "electron-only" detection strategies to ameliorate the problems associated with the poor hole mobilities and short hole trapping times in CZT. Some one uses the small-pixel-effect so that charge generated inside the detector induces pixel currents only briefly before impinging on the anode pixels. The anode signals become largely independent of the location of the charge generation and the "depth of interaction" (DOI). In additional, DOI correction can be used to further improve the energy resolution. The DOI can be estimated from measuring the drift time of the electrons and from the anode to cathode charge signal ratio

[1,30-31].

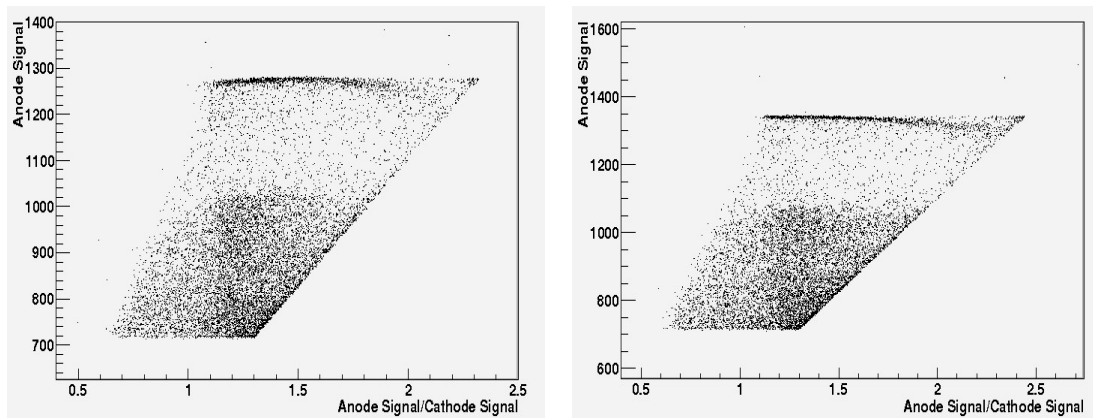

Figure 9: The two panels show the anode to anode-to-cathode ratio correlation for two detectors illuminating by $^{137}$Cs source, the 0.75×2×2 cm$^3$ detector from the company Orbotech (left panel), and the 1×2×2 cm$^3$ detector from the company eV-Products (right panel). The anode signal depends only weakly on the DOI.

Figure 9 shows the correlation of the anode signal to the anode to cathode ratio for two detectors, the 0.75×2×2 cm$^3$ detector from the company Orbotech, and the 1×2×2 cm$^3$ detector from the company eV-Products. For both detectors, only a weak dependence of the anode signal on the DOI can be recognized. Figure 10 shows the energy spectra before and after correction of the anode signal for the DOI illuminating by $^{137}$Cs source. The energy resolution increases slightly. After correction, the best pixels for the two detectors improve in energy resolution from 1.01% (6.69 keV) to 0.79% (5.23 keV) and 0.89% (5.89 keV) to 0.61% (4.04 keV), respectively.

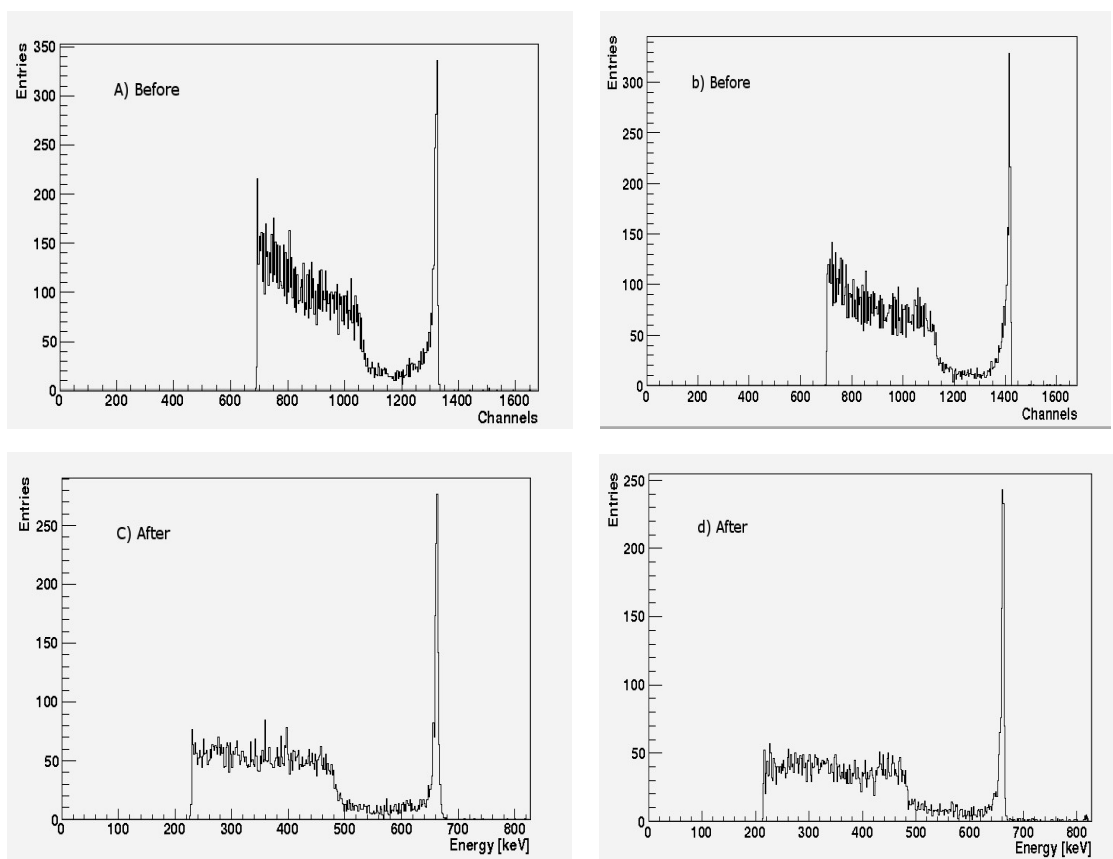

Figure 10: The two panels show $^{137}$Cs energy spectra measured before and after correction for the depth of interaction for the 0.75×2×2 cm$^3$ detector (a,c) from the company Orbotech (left panel), and the 1×2×2 cm$^3$ detector (b, d) from the company eV-Product (right panel) illuminating by $^{137}$Cs source.

## 4.3 Reproducibility of Results

We investigated the reproducibility of our results with a 0.5×2×2 cm$^3$ detector from the company Orbotech. For this detector, we performed five iterations. Each iteration included the removal of previously deposited pixels, the new deposition of 8×8 Ti pixels (including a polish with abrasive paper and Al suspension, wet etching, and contact deposition with the e-beam as describes in Sect. 2), and tests of the detector with $^{57}$Co and $^{137}$Cs sources. The five detector fabrication runs gave highly consistent results. Averaged over all the pixels of the detector, the five runs gave 122 keV energy resolutions of 3.46 keV, 3.46 keV, 3.28 keV, 3.35 keV, and 3.44 keV. For one of the best pixels close to the center of the detector, we obtained energy resolutions of 2.6 keV, 2.6 keV, 2.7 keV, 2.7 keV, and 2.8 keV. Figure 11 shows an overlay of the $^{57}$Co energy spectra averaged over the central 6×6 pixels of the detector for the five fabrication runs. The figure shows that the shape of the energy spectra (including the peak to valley ratios) did not change from run to run. The five runs gave 662 keV energy resolutions of 8.44 keV, 9.3 keV, 9.25 keV, 9.8 keV, and 11.0 keV. Although the trend is not highly significant, the energy resolutions might get slightly worse from run to run. This result may be explained in part by the detector thickness decreasing from 0.50 cm at the start of the study to 0.45 cm after the fifth polish. The smaller detector thickness results in a smaller aspect ratio of pixel pitch to detector thickness. The effect may be more pronounced at 662 keV than at 122 keV owing to two reasons: (i) the results at 662 keV are less impacted by electronic noise than the results at 122 keV; any systematic effect will thus show up more at 662 keV than at 122 keV; (ii) the 122 keV photons interact more closely to the cathode than the 662 keV events; for 122 keV events, a small pixel effect is thus less important than for 662 keV events. The relatively poor 662 keV energy resolutions for the smallest pixel pitch in Fig. 5 may be explained - in part - by the fact that we started the test series with larger pixel pitches and proceeded towards smaller pixel pitches. However, the effect from the thinning of the detectors should be rather small as only three fabrication steps were involved in the study; furthermore, the effect should even be less important for the 0.75 cm and 1 cm thick detectors in Fig. 5 than for 0.5 cm thick detectors as used for the reproducibility studies described in this section.

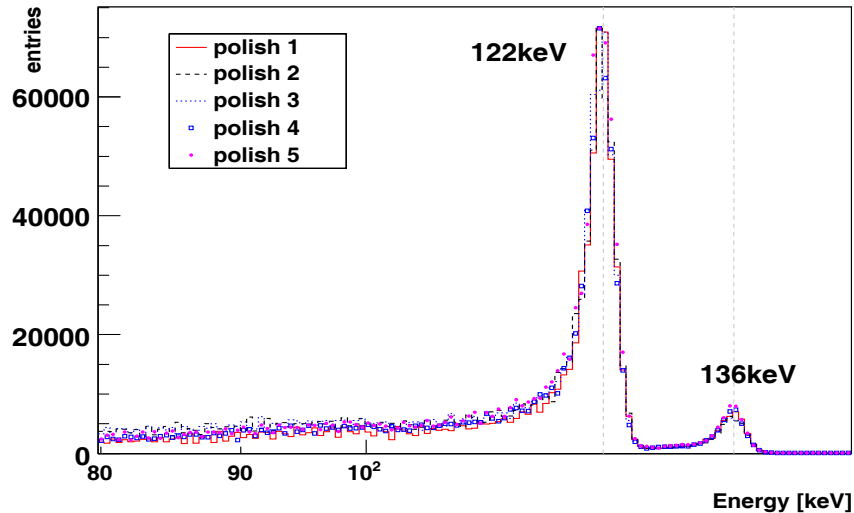

Figure 11: Results from a reproducibility study. We performed five iterations of polishing off previously deposited contacts, depositing 8×8 pixels (pixel pitch of 2.5 mm), and taking $^{57}$Co and $^{137}$Cs energy spectra. The figure shows the $^{57}$Co energy spectra obtained after each of the five polishes. The five iterations gave highly consistent results – even though the detector thickness decreased from 0.50 cm at the beginning of the study to 0.45 cm after the fifth polish. Note that the energy spectra shown here were obtained with the central 6×6 pixels which perform slightly better than the outer ring of 28 pixels.

## 5. Discussion

Using the NCI ASIC developed at the Brookhaven National Laboratory, we have developed a readout system for the test of CZT detectors. The system achieves a low readout noise of about 2 keV FWHM. We have used the system to test a number of MHB and HPB CZT crystals of different thicknesses contacted with different pixel patterns. Our main result is that all the tested detectors give excellent energy resolutions - even without any depth of interaction correction.

We reported excellent results obtained with both, MHB and HPB material. Unfortunately, our study did not reveal clear trends of the energy resolution as function of the detector thickness or pixel pitch. We are continuing the detector fabrication and evaluation program. The current work focuses on the evaluation of large volume ($0.5 \times 3.9 \times 3.9$ cm$^3$) CZT detectors from the company Orbotech, the development of CZT detectors with cross-strip readout, and the test of detectors with small pixel pitches of 350 μm and 600 μm, and the development and test of a scalable readout system for arrays of CZT detectors

**Acknowledgements**

This work is supported by NASA under contract NNX07AH37G and NNX10AJ56G, and the DHS under contract 2007DN077ER0002. We thank electrical engineer R. Bose and technicians G. Simburger and D. Braun for their support.